\documentclass[twocolumn,preprintnumbers,amsmath,amssymb,prb]{revtex4}
\usepackage[english]{babel}
\usepackage{graphicx}
\usepackage{dcolumn}
\usepackage{bm}
\usepackage[version=3]{mhchem}

\begin{document}


\title{First-Principle Study of Phosphine Adsorption on Si(001)-2$\times$1-Cl}

\author{Tatiana V. Pavlova$^{1,2}$}
\email{pavlova@kapella.gpi.ru}
\author{Georgy M. Zhidomirov$^{1,2}$}
\author{Konstantin N. Eltsov$^{1,2}$}

\affiliation{$^{1}$Prokhorov General Physics Institute of the Russian Academy of Sciences, Moscow, Russia}
\affiliation{$^{2}$National Research University Higher School of Economics, Moscow, Russia}


\begin{abstract}

This paper presents a DFT study for phosphine adsorption on a
Si(001)-2$\times$1 surface covered by a chlorine monolayer,
including adsorption on local defects, i.e. mono- and bivacancies in
the adsorbate layer (Cl, Cl$_2$), and combined vacancies with
removed silicon atoms (SiCl, SiCl$_2$). Activation barriers were
found for the adsorbing PH$_3$ to dissociate into PH$_2$+H and
PH+H$_2$ fragments; it was also established that phosphine
dissociation on combined vacancies is possible at room temperature.
If there is a silicon vacancy on the surface, phosphorus settles in
the Si(001) lattice as PH (if the vacancy is SiCl) or as PH$_2$  (if
the vacancy is SiCl$_2$). This paper suggests a method to plant a
separate phosphorus atom into the silicon surface layer with atomic
precision, using phosphine adsorption on defects specially created
on a Si(001)-2$\times$1-Cl surface with an STM tip.


\end{abstract}

\maketitle

\section{Introduction}
Developing atomic-scale nanoelectronic components, in  particular
building blocs for the  quantum computer on nuclear spins of
phosphorus in Si \cite{bib:Kane}, requires methods for incorporating
single impurity atoms into silicon. STM lithography
--- to date, the most precise method of manipulating a solid surface
--- offers such a possibility by creating or deleting atomic
structures with pinpoint impacts of a scanning tunneling microscope
(STM) tip. In Ref. \cite{bib:STM-H}, a chemisorbed hydrogen
monolayer is used as the resist for a Si(001)-2$\times$1 surface.
First the mask is prepared by electron-stimulated desorption of
selected hydrogen atoms with an STM tip, then phosphine (PH$_3$) is
adsorbed on the masked surface, with PH$_3$ adsorbing only to places
free of hydrogen. As this surface is heated to 350\,$^{\circ}$C, the
phosphine molecule dissociates, and a phosphorus atom is
incorporated into the Si(001)-2$\times$1 atomic lattice.

Chlorine monolayer on a silicon surface may also be considered as a
resist for STM lithography. The difference between the chlorine and
hydrogen monolayers is most pronounced in thermal desorption
experiments: SiCl$_2$ compound removes from the
Si(001)-2$\times$1-Cl surface \cite{bib:TPD_ClSi, bib:TPD_ClSi1},
while hydrogen desorbs from Si(001)-2$\times$1-H as H$_2$ molecule
\cite{bib:TPD_H}. Therefore, one can expect that as a result of
electron-stimulated desorption of chlorine atom from monolayer, atomic defects
containing a silicon vacancy will be created by STM tip.

The aim of the work is to study PH$_3$ interaction with
Si(001)-2$\times$1-Cl surface (with and without vacancies) in order
to predict the possibility to utilize chlorine resist for phosphorus
incorporation into silicon lattice. Phosphine adsorption on clean
Si(001) surface is a broadly discussed (see Ref. \cite{bib:PH3} and
Refs. therein) and well-researched topic. Adsorption of phosphine on
hydrogenated Si(001) surface with defects was considered for
studying the incorporation of phosphorus atom into the silicon
lattice \cite{bib:STM-H}. However, no effort has to our knowledge
been spent so far on phosphine interaction with a Si(001) surface
whose electron structure was modified by chemisorbed chlorine.

This paper presents the results of theoretical inquiries into PH$_3$
adsorption on an ideal chlorinated Si(001)-2$\times$1 surface and
the same surface with local defects --- Cl, Cl$_2$, SiCl, and
SiCl$_2$ vacancies. We have created DFT models of atomic structures
containing a PH$_3$ molecule and PH$_2$ and PH dissociation
fragments; established the most beneficial atomic configurations for
each of these defects; mapped the possible ways of PH$_3$
dissociation; and found the corresponding activation barriers. On
the basis of this data, we discuss the process of implanting a
phosphorus atom into a silicon lattice as a result of phosphine
adsorption on atomic defects in Si(001)-2$\times$1-Cl.

\section{Method}

Spin polarized calculations were made on the basis of the density
functional theory (DFT) as provided by the VASP software package
\cite{bib:Kresse1993,bib:Kresse1996}. We used generalized gradient
approximation (GGA), PBE exchange correlation functional
\cite{bib:PBE}, and Grimme corrections to van der Waals interactions
\cite{bib:D2}. The Si(001)-2$\times$1 surface was modeled by
periodic 4$\times$4 cells. Each cell contained eight silicon atomic
layers, the three bottom layers are fixed. Chlorine atoms and a
phosphine molecule were placed on the top of the slab, while the
bottom surface was covered with hydrogen. The slabs were separated
by a vacuum spacing of 15\,{\AA}.

The PH$_3$ adsorption energy was calculated as the difference
between the total energy of a chlorinated surface
Si(001)-2$\times$1-Cl with the adsorbate ($E_{PH_3+surf}$) and the
total energies of the Si(001)-2$\times$1-Cl surface ($E_{surf}$) and
a PH$_3$ molecule in the gaseous phase ($E_{PH_3}$):
\begin{equation}
E_{a} = E_{PH_3+surf} - E_{surf} - E_{PH_3}. \label{eq:1}
\end{equation}
The underlying assumption was that the atomic defects had been
created on the Si(001)-2$\times$1-Cl before the PH$_3$ adsorption
(e.g., with an STM tip), therefore equation (\ref{eq:1}) did not
include the energy spent on creating them.

Activation barriers ($E_{act}$) were calculated using the NEB
(nudged elastic band) method \cite{bib:NEB} provided within VASP.

\section{Results and Discussion}

Since there are no studies of phosphine adsorption on chlorinated
silicon surface, we tested our calculation method on a well-studied
PH$_3$/Si(001)-c(4$\times$2) system \cite{bib:PH3}. The ground state
of an atomically clean Si(001) surface is the c(4$\times$2)
reconstruction, in which dimers in neighboring rows are tilted in
opposite directions \cite{bib:c4x2}. Our values for Si interatomic
distances in a dimer (2.36\,{\AA}) and the dimer tilt angle
(18$^{\circ}$) on a clean Si(001)-c(4$\times$2) surface are in good
agreement with experimental data \cite{bib:surf_int_book} and other
calculations (see, e.g., Ref. \cite{bib:NH3}). It is known that at
room temperature, PH$_3$, as it adsorbs on a Si(001), dissociates
into PH$_2$ + H in the first stage of the reaction \cite{bib:PH3},
with the PH$_2$ + H adsorption energy on Si(001) calculated
theoretically somewhere between $-1.88$\,eV and $-2.29$\,eV (see
ref. \cite{bib:PH3} and Refs. therein). Our calculated value $E_a =
-2.03$\,eV is in good agreement with previous calculations and with
an experimental estimate of $-1.8$\,eV made in Ref. \cite{bib:PH3}
on the basis of thermal-desorption spectroscopy data
\cite{bib:TPD_PH2}. Distance $d$ from a P atom to the nearest Si
atom was 2.28\,{\AA}, from a dissociated H atom to the Si atom
$d$(Si-H) = 1.50\,{\AA}, from a P atom to the H atom in the PH$_2$
fragment $d$=1.44\,{\AA} ($d$(P-H) = 1.43\,{\AA} in a free PH$_3$
molecule).

\subsection{\label{sec:ads}PH$_3$ adsorption on Si(001)-2$\times$1-Cl}

After a Si(001) surface is chlorinated, the c(4$\times$2)
reconstruction is partially removed and changed into the
(2$\times$1) reconstruction. Our calculated value for the bond
length $d$(Si-Cl) is 2.07\,{\AA}, which is in agreement with both
experimental \cite{bib:surf_int_book} and theoretical
\cite{bib:NH3,bib:Cl_Br_Si} data. The distance between the Si atoms
in a dimer increases by 0.07\,{\AA} after chlorine adsorption
($d$(Si--Si) = 2.43\,{\AA}). We also calculated the energies and
activation barriers for phosphine adsorption and dissociation for an
ideal (defect-free) Si(001)-2$\times$1-Cl surface and for the same
surface with vacancy defects Cl, Cl$_2$, SiCl and SiCl$_2$. For the
ideal surface, we looked at two PH$_3$ adsorption scenarios: the
molecule adsorbs on top of or under the chlorine layer. For a
defected surface (Cl, Cl$_2$, SiCl, SiCl$_2$ vacancies), we looked
at PH$_3$ adsorption and dissociation in all possible scenarios, of
which only those were included in the paper whose final state is
more preferable than the initial state.

\subsubsection{Defect-free Si(001)-2$\times$1-Cl}

\begin{figure}
\centering
\includegraphics[width=8cm]{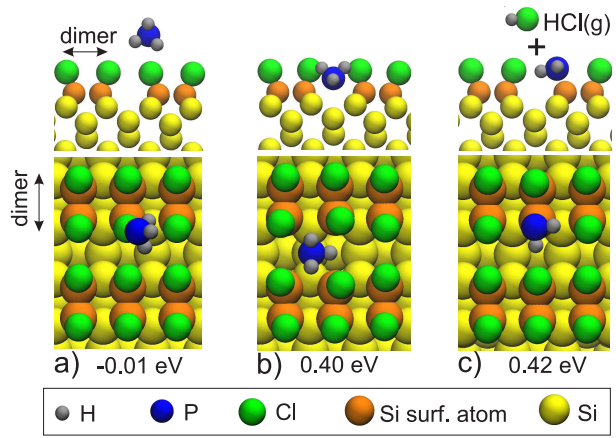}
\caption{\label{figNovac}Structures (top view and side view) and
adsorption energies of PH$_3$ on Si(001)-2$\times$1-Cl: (a) PH$_3$
on top of the chlorine layer, (b) PH$_3$ under the chlorine layer,
(c) substitution PH$_2$ instead of Cl with HCl molecule formation in
gas phase.}
\end{figure}

The adsorbed PH$_3$ molecule remains close to the surface
(Fig.~\ref{figNovac}a). There is almost no interaction with chlorine
($E_a = -0.01$\,eV), and the distance from the P atom to the nearest
Cl atom is 3.63\,{\AA}. The molecule is adsorbed without any
activation barrier and does not dissociate.

A study of interaction between ammonia (NH$_3$) and a
Si(001)-2$\times$1 chlorinated surface \cite{bib:NH3} demonstrated
that embedding of an NH$_2$ fragment instead of a chlorine atom is
possible (with formation of HCl molecule) if NH$_3$ is incorporated
under the chlorine layer. Our calculated energy values for PH$_3$
on-top adsorption ($-0.01$\,eV, Fig.~\ref{figNovac}a), PH$_3$ adsorption
under the chlorine layer between Si dimer rows ($0.40$\,eV, Fig.~\ref{figNovac}b), and the adsorption of a PH$_2$ fragment bound to a
Si atom with a HCl molecule close to the surface ($0.42$\,eV, Fig.~\ref{figNovac}c), are close to values reported in Ref.
\cite{bib:NH3} for the energies of initial ($-0.02$\,eV),
transitional ($0.40$\,eV) and final ($0.45$\,eV) states of
substituting NH$_2$ instead of a Cl atom with HCl molecule
formation. This suggests that the transitional state energies of
NH$_3$ and PH$_3$ on Si(001)-2$\times$1-Cl may also be approximately
the same. Our calculated initial, transitional, and final state
PH$_3$ adsorption energies, and activation barrier data from Ref.
\cite{bib:NH3}, bring us to a conclusion that a PH$_2$ fragment can
only be placed on Si(001)-2$\times$1-Cl instead of a Cl atom if it
overcomes the activation barrier of $\approx 1.6$\,eV. Thus,
reaction of substitution a PH$_2$ fragment instead of a Cl atom on
an ideal Si(001)-2$\times$1-Cl surface is endothermic ($E_a > 0$ in
the final state) and unfavorable.

\subsubsection{Si(001)-2$\times$1-Cl with a single vacancy in a chlorine monolayer}

\begin{figure}
\centering
\includegraphics[width=6.5cm]{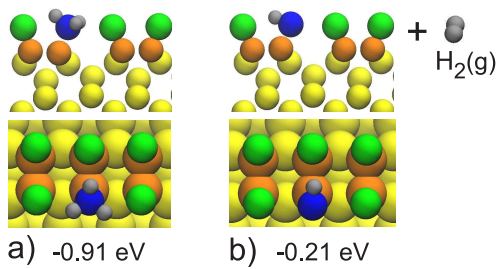}
\caption{\label{figCl}Structures (top view and side view) and
adsorption energies of phosphine and its fragments into a single
vacancy in chlorine monolayer on Si(001)-2$\times$1-Cl: (a) PH$_3$
adsorption; (b) PH + H$_2$ (g) adsorption.}
\end{figure}

As a phosphine molecule is adsorbed into a single vacancy in a
chlorine monolayer on Si(001)-2$\times$1-Cl, PH$_3$ takes the place
vacated by a chlorine atom (Fig.~\ref{figCl}a). There is no activation
barrier for phosphine adsorption. The calculated Si-P bond length is
2.31\,{\AA}, distance $d$(P--H) = 1.41\,{\AA}.

Phosphine, dissociating in a single chlorine vacancy to produce a PH
fragment, also necessarily produces an H$_2$ molecule (Fig.~\ref{figCl}b)
because there are no more chlorine-free places on the surface. This
PH + H$_2$ dissociation is unfavorable because the adsorption energy
of this system ($E_a = -0.21$\,eV) is smaller than the energy of the
adsorbed PH$_3$ molecule ($E_a = -0.91$\,eV).

\subsubsection{Si(001)-2$\times$1-Cl with a double vacancy in a chlorine monolayer}

\begin{figure}
\centering
\includegraphics[width=8cm]{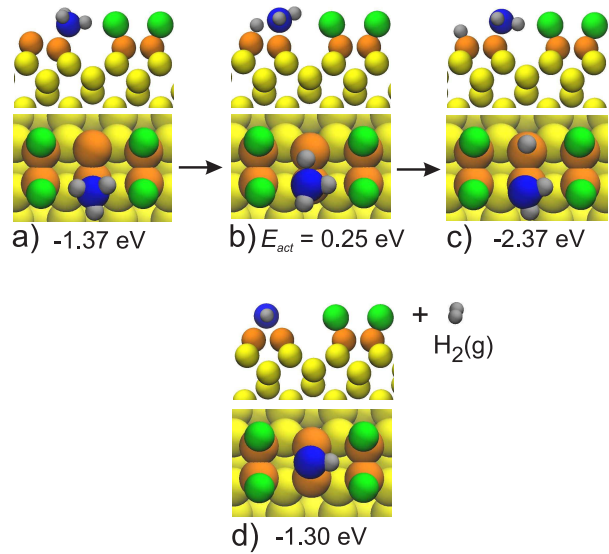}
\caption{\label{figCl2}Structures (top view and side view) and
adsorption energies of phosphine and its fragments into a double
vacancy in chlorine monolayer on Si(001)-2$\times$1-Cl: (a) PH$_3$
adsorbed molecule; (b) intermediate state in the process of PH$_3$
dissociation into PH$_2$ + H (c); (d) PH fragment in the center of a
Si-Si dimer, and H$_2$ molecule in gas phase.}
\end{figure}

Figure \ref{figCl2}a shows a PH$_3$ molecule adsorbed on a Si-Si dimer free
of chlorine atoms. A phosphorus atom is bound to one of the silicon
atoms. The Si-P bond length is 2.30\,{\AA} ($d$(P-H) = 1.41\,{\AA});
the adsorption energy is $-1.37$\,eV; there is no activation barrier
(Fig.~\ref{figCl2}a). However, a phosphine molecule in a double chlorine
vacancy can dissociate into PH$_2$ and H fragments and thus create
the most energetically favorable state ($E_a = -2.37$\,eV, Fig.~\ref{figCl2}c). The dissociation process goes through an
intermediate state shown in Fig.~\ref{figCl2}b and requires an activation
energy of $0.25$\,eV. While on a clean Si(001)-c(4$\times$2)
surface, PH$_2$ dissociates further into PH or P \cite{bib:PH3}, it
is not preferable for it to do so on a chlorinated surface because
there are no unoccupied space for the hydrogen atoms. Nonetheless,
there is yet another state with a negative adsorption energy
(Fig.~\ref{figCl2}d), in which a PH fragment occupies a bridge position
over the Si-Si dimer ($d$(Si-P) = 2.29\,{\AA}), while the remaining
hydrogen atoms form an H$_2$ molecule. This state ($E_a =-1.30$\,eV)
is less preferential than phosphine molecule adsorption ($E_a
=-1.37$\,eV). From this adsorption site, the phosphorus atom can (in
principle) migrate to the position between two silicon dimers, so
that the P atom is bound to three Si atoms (one Si atom of each
dimer and one Si atom of underlying layer) and H atom remains on the free
silicon atom. Note, this position is the most favorable on a clean
silicon surface \cite{bib:PH3}. However, in the presence of only one
chlorine-free Si dimer, this state is less favorable than those shown
in Fig.~\ref{figCl2}, because one of the Si atoms bound to P is
also bound to one chlorine atom. This is in agreement with
the finding \cite{bib:PH3} that three free dimers on a
Si(001)-c(4$\times$2) surface are required for full phosphine
dissociation.

\subsubsection{Si(001)-2$\times$1-Cl with a combined SiCl vacancy}

\begin{figure}
\centering
\includegraphics[width=8cm]{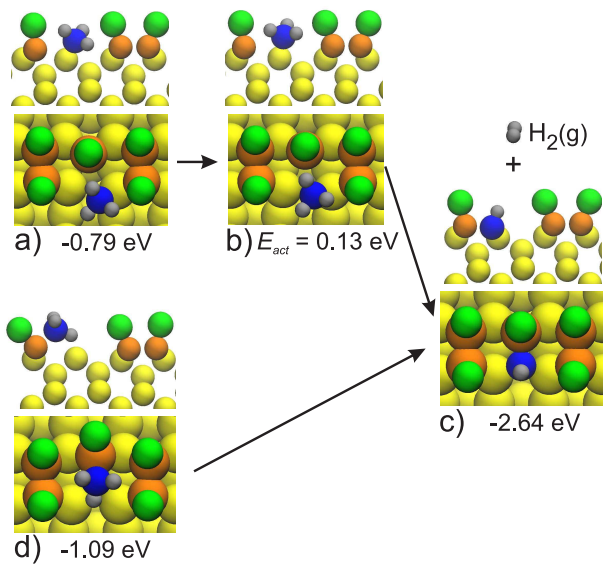}
\caption{\label{figSiCl}Structures (top view and side view) and
adsorption energies of phosphine and its fragments into a combined
SiCl vacancy on Si(001)-2$\times$1-Cl: (a) PH$_3$ molecule adsorbed
on lower-layer silicon atom; (b) intermediate state in the process
of PH$_3$ dissociation into PH + H$_2$ (c); (d) PH$_3$ molecule
adsorbed on the remaining silicon atom of a broken dimer.}
\end{figure}

In phosphine adsorption into combined vacancies in which both
silicon and chlorine have been removed, a phosphine molecule
approaching such a Si(001)-2$\times$1-Cl surface may create a single
bond either with a lower-layer silicon atom (Fig.~\ref{figSiCl}a) or with
the remaining silicon atom of a broken dimer (Fig.~\ref{figSiCl}d)
because PH$_3$ can not create bonds to two Si atoms. There is no
activation barrier in both cases but the latter option is more
preferable (its $E_a = -1.09$\,eV against the former option's $E_a =
-0.79$\,eV). Bonded to the top Si atom (Fig.~\ref{figSiCl}d), the
phosphine molecule dissociates into PH + H$_2$ (Fig.~\ref{figSiCl}c) with
almost no activation barrier ($E_{act} = 0.02$\,eV). If the
phosphine molecule is bonded to a lower-layer silicon atom
(Fig.~\ref{figSiCl}a), its dissociation results in the same final state
(Fig.~\ref{figSiCl}c) but only over an activation barrier of $0.13$\,eV
(Fig.~\ref{figSiCl}b). The structure in which the PH fragment is
incorporated into the silicon lattice and an H$_2$ molecule is
formed (Fig.~\ref{figSiCl}c) is the most preferable ($E_a = -2.64$\,eV,
phosphorus bond lengths to two lower-layer silicon atoms are
2.32\,{\AA}, to the dimer Si atom 2.37\,{\AA}, to the hydrogen atom
1.42\,{\AA}).

\subsubsection{Si(001)-2$\times$1-Cl with a combined SiCl$_2$ vacancy}

\begin{figure}
\centering
\includegraphics[width=8cm]{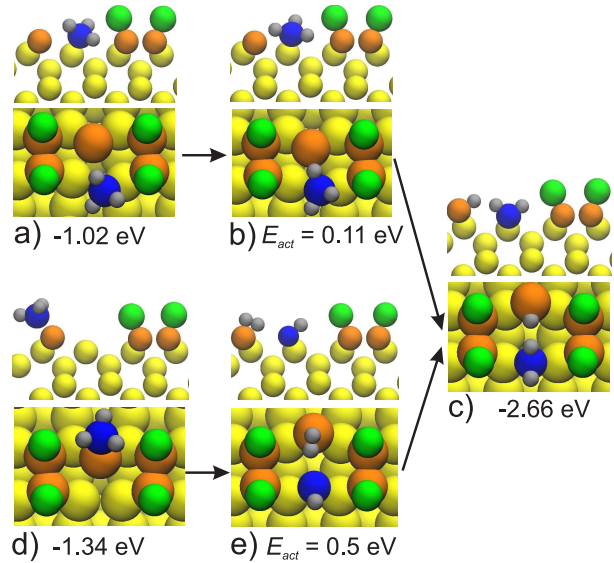}
\caption{\label{figSiCl2} Structures (top view and side view) and
adsorption energies of phosphine and its fragments into a combined
SiCl$_2$ vacancy on Si(001)-2$\times$1-Cl: (a) PH$_3$ molecule bound
to a lower-layer silicon atom; (b) intermediate state between (a)
and (c); (c) the most preferable state after PH$_3$ dissociation
into PH$_2$ and H; (d) PH$_3$ molecule bound to a broken-bond dimer
silicon atom; (e) intermediate state between (d) and (c).}
\end{figure}

As for a SiCl vacancy, the phosphine molecule is adsorbed into a
SiCl$_2$ vacancy either through bonding to a lower-layer silicon
atom (Fig.~\ref{figSiCl2}a) or occupying a position close to a silicon
atom in broken dimer (Fig.~\ref{figSiCl2}d). Neither process has an
activation barrier. A PH$_3$ molecule in a SiCl$_2$ vacancy can
transfer one hydrogen atom to the broken-bond silicon atom, while
the remaining PH$_2$ fragment can meanwhile form two bonds to
silicon atoms from the lower layer (Fig.~\ref{figSiCl2}c). The adsorption
energy is minimal in this state ($E_a = -2.66$\,eV), the Si-P bond
lengths are \,{\AA} ($d$(P-H) = 1.42\,{\AA}). PH$_3$ dissociation
into PH$_2$ +H from positions shown in Fig.~\ref{figSiCl2}a and
Fig.~\ref{figSiCl2}d requires activation energies of $0.11$\,eV (Fig.~\ref{figSiCl2}b) and
$0.50$\,eV (Fig.~\ref{figSiCl2}e), respectively. Note that the reaction paths with the HCl
molecule formation are less favorable than the pathways with the
H$_2$ formation for the Si(001)-2$\times$1-Cl surface with Cl,
Cl$_2$, and SiCl vacancies.

\subsection{\label{sec:inc}Incorporating a phosphorus atom into the Si(001) lattice}

If hydrogen resist is used on Si(001), it is only possible to plant
a phosphorus atom on a Si(001) surface with an error of several
lattice spacings \cite{bib:D-PRL} because three free silicon dimers
are required for PH$_3$ dissociation \cite{bib:PH3}, and as the
surface is heated up, any of the neighboring substrate atoms may be
substituted by phosphorus atom. This precision is sufficient to
create a single electron transistor to read the qubit state but not
to create a network of qubits, where phosphorus atoms must be
incorporated exactly into the positions of specifically selected
silicon atoms. In Refs. \cite{bib:J_osc_Koiller,bib:J_osc}, it was
demonstrated that, if two neighboring phosphorus atoms are not
located precisely along high-symmetry directions on a silicon
surface (diverging from that line by one or several lattice
spacings), the exchange interaction between them oscillated and
decreased exponentially, which is a serious problem in the way of
creating a two-qubit register of a quantum computer.

We suggest a procedure for incorporating phosphorus atoms into the
top Si(001)-2$\times$1 layer with atomic precision. It is based on
two operations. The first one is STM lithography on Si(001) surface
with a chlorine monolayer used as the resist to create atomic
defects containing a silicon vacancy. We believe that there is a
controlled way to create vacancies with silicon atoms for
Si(001)-2$\times$1-Cl surfaces. The second operation is phosphine
adsorption on defects containing silicon vacancies. According to our
calculations summarized below, phosphine adsorption on such defects
results in incorporating a phosphorus atom exactly into the site
previously occupied by a silicon atom removed with an STM tip.

As phosphine is adsorbed onto a perfect Si(001)-2$\times$1-Cl
surface, there is little interaction between the gas and the
chlorine monolayer because the PH$_3$ adsorption energy is close to
zero ($-0.01$eV). A phosphine molecule dissociation, so that the
PH$_2$ fragment is put into the place of a chlorine atom and an HCl
molecule is formed, is endothermic process and requires an
activation energy of about $\sim 1.6$\,eV, which makes this process
unfavorable at room temperature.

If there is a single vacancy in the chlorine monolayer, the
phosphine molecule settles into the vacant place and does not
dissociate. If there is a double vacancy in the chlorine monolayer,
i.e. the phosphine molecule is adsorbed onto a free Si-Si dimer, it
dissociates into PH$_2$ and H at room temperature (the activation
energy is $0.25$\,eV). In this case P atom incorporation into
silicon surface layer requires two additional Si dimers free from Cl
atoms and heating to 350\,$^{\circ}$C \cite{bib:PH3}.

At room temperature, phosphine adsorption onto a
Si(001)-2$\times$1-Cl surface with combined vacancies (SiCl or
SiCl$_2$) results in phosphine dissociation, with the fragments
taking the most favorable positions: PH in the SiCl vacancy, PH$_2$
+ H in the SiCl$_2$ vacancy. The adsorption into the SiCl and
SiCl$_2$ vacancies ($E_a =-2.64$\,eV and $E_a =-2.66$\,eV,
respectively) are more preferable than into the Cl and Cl$_2$
vacancies ($E_a =-0.91$\,eV and $E_a =-2.37$\,eV, respectively).
Thus, it is possible to incorporate a phosphorus atom into a silicon
lattice at room temperature as part of a PH fragment adsorbed on a
SiCl vacancy or PH$_2$ fragment adsorbed on a SiCl$_2$ vacancy
without undesirable heating, which is not possible if chemisorbed
hydrogen is used as the resist \cite{bib:STM-H}.

The resulting surface is Si(100) under a chlorine monolayer, with
phosphorus atoms embedded in the silicon lattice. One or two
hydrogen atoms may be bound with each phosphorus atom. To use these
phosphorus atoms as quantum computer elements (such as registers,
wiring, or convertors), chlorine and hydrogen should be removed, and
a sufficiently thick crystalline silicon layer (25 to 50 nm) should
be grown on top of the surface enough to minimize the impact of the
surface on computing operations.

\section{\label{sec:conc}Conclusions}

On the basis of DFT modeling, we studied phosphine adsorption on
Si(001)-2$\times$1-Cl surface, including surface with Cl, Cl$_2$,
SiCl, and SiCl$_2$ vacancies. We have demonstrated that it is
possible to incorporate a phosphorus atom into the silicon lattice
if phosphine is adsorbed into defects with a silicon vacancy. We
found that phosphorus is best positioned to occupy the vacant place
in the lattice as part of a PH fragment in a SiCl vacancy
(Fig.~\ref{figSiCl}c) or as part of a PH$_2$ fragment in a SiCl$_2$
vacancy (Fig.~\ref{figSiCl2}c). Phosphine molecules adsorbed on a
chlorinated defect-free Si(001)-2$\times$1 surface are not bound to
silicon.

Our calculations confirm that it is possible to incorporate a
phosphorus atom precisely into a specific place of the silicon
lattice on Si(001)-2$\times$1 through a chemisorbed chlorine
monolayer mask. As we have demonstrated, operationally this requires
a specific type of vacancies on the Si(001)-2$\times$1-Cl surface,
and we expect STM lithography to create specific types of defects to
be developed.

\section{Acknowledgements} This study was supported by the Russian Science
Foundation (Grant 16-12-00050). We also thank the Joint
Supercomputer Center of RAS for providing the necessary computing
power.


\end{document}